\documentclass[a4paper, 11pt]{article}
\usepackage{ichmt}
\usepackage{subfigure} 
\usepackage{indentfirst}
\usepackage{setspace}
\usepackage{chngpage}
\usepackage[square]{natbib} 
\usepackage{graphicx,amssymb,verbatim} 

\date{}
\begin{document} 
\changetext{-3cm}{}{}{}{}
\title{Integral analysis of laminar indirect free convection
boundary layers with weak blowing for \mbox{Schmidt no. $\sim$ 1}} 

\author{Baburaj A.Puthenveettil and Jaywant H.Arakeri\\ Department of
  Mechanical Engineering,\\ Indian Institute of Science, Bangalore
  560012\\(jaywant@mecheng.iisc.ernet.in)}
\maketitle
\begin{abstract}
  Laminar natural convection at unity Schmidt number over a horizontal
  surface with a weak normal velocity at the wall is studied using an
  integral analysis. To characterise the strength of the blowing, we
  define a non-dimensional parameter called the blowing parameter.
  After benchmarking with the no blowing case, the effect of the
  blowing parameter on boundary layer thickness, velocity and
  concentration profiles is obtained. Weak blowing is seen to increase
  the wall shear stress.  For blowing parameters greater than unity,
  the diffusional flux at the wall becomes negligible and the flux is
  almost entirely due to the blowing.
\end{abstract} 
\onehalfspacing
\section*{\underline{Introduction}}
\label{sec:introduction}
Fluid motion driven by an indirectly induced pressure gradient normal
to the direction of the density potential difference is termed as
indirect natural convection ~\citep{geb,blt}. An example is natural
convection occurring over a horizontal heated surface. At large
Grashoff numbers, the flow is of boundary layer type , and can be
described by similarity solutions~\citep{gzc,rc}.  Such boundary
layers could form over a porous surface with  blowing at the surface,
the effect of which on the behavior of the boundary layer is not
obvious.
\changetext{3cm}{}{}{}{}
Previous investigations on the effect of blowing on indirect free
convection boundary layers were conducted by Clarke \& Riley
~\citep{cr} and Gill et.al.~\citep{gzc} using similarity analysis of
the boundary layer equations.  The boundary layer equations admit
similarity solutions only for a restrictive and interdependent spatial
distribution of wall velocity and density potential. Hence, the scope
of these solutions is limited.  We study indirect natural convection
boundary layers for unity Schmidt(or Prandtl) number fluids, when a
weak normal velocity is imposed at the wall.  The wall velocity and
density potential are not functions of horizontal distance. The
imposed velocities are considered to be weak in the sense that they
are of the order of vertical characteristic velocity in the boundary
layer when no blowing is present. An integral method is used to obtain
approximate solutions to the boundary layer equations for blowing
velocities and wall concentrations that are constant in the horizontal
direction.

The notation used is plain symbols for nondimensional variables,
superscript $\,\tilde{}\,$ for dimensional variables and subscript c
for characteristic scales.  The paper is organised in the following
manner.  Using the characteristic scales in the problem,the boundary
layer equations are derived. Integral equations are formulated by
integrating the boundary layer equations across the boundary layer
thickness, using profiles that satisfy the boundary conditions.  The
solution procedure of the resulting initial value problem is explained
and the results are initially compared with the similarity solutions
of Rotem and Classen~\citep{rc} for no blowing case. The effect of
blowing on the boundary layer thickness, concentration and velocity
profiles, and mass flux are studied.
\section*{\underline{Boundary layer approximation}}
\label{sec:bound-layer-appr}
\begin{figure}[hbp]
\centering
  \includegraphics[width=0.45\textwidth]{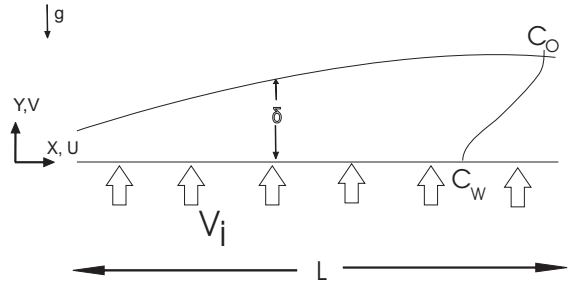}
\caption{Schematic of the problem studied }\label{sche}
\end{figure}
Consider a fluid having a concentration $\tilde{C}_o$ of some species
above a horizontal porous surface maintained at $\tilde{C}_w$, with
$\tilde{C}_w \,<\,\tilde{C}_o$.  Also, a fluid with concentration
$\tilde{C}_w$ flows through the surface with a normal velocity
$\tilde{v}_w$. (The problem is equivalent to natural convection over a
heated plate with blowing, with $\tilde{C}_w$ representing the wall
temperature and $\tilde{C}_o$ the temperature of the ambient.) The
geometry of the problem studied along with the variables used are
shown in Figure~(\ref{sche}).  $\tilde{u},\,\tilde{v}$ are the
velocity components in the $\tilde x , \tilde y$ directions, and the
boundary layer thickness is $\tilde\delta(\tilde x)$.The wall velocity
is considered to be independent of the density potential.

Let the characteristic length scales in the horizontal and vertical
directions be $x_{c}=L,$ and $y_{c}=\delta_c,$ a characteristic
boundary layer thickness.  L is taken as the horizontal dimension of
the surface.  Let $\,u_{c}\,\mathrm{and}\,v_{c}$ denote the
characteristic horizontal and vertical velocities, $p_{c}$ the
characteristic pressure and $c_{c}=\tilde C_o-\tilde C_{w}=\Delta
\tilde C$ the characteristic concentration difference.

Following Rotem and Classen~\citep{rc}, for $Sc\sim 1$
\begin{equation}\label{dcucpc} 
\delta_c\sim\frac{L}{Gr_L^{1/5}}\,,\;\;u_c\sim\frac{\nu}{L}Gr_L^{2/5}\,,\;\;
v_c\sim\frac{\nu}{L}Gr_L^{1/5},\;\;
p_c\sim\rho(\nu/L)^2Gr_L^{4/5}
\end{equation}
where, $Gr_ L\,=\frac{g\beta\Delta \tilde{C} L^3}{\nu^2}$, the
Grashoff number.  Here $g\,,\beta\,,\nu\,$ denote acceleration due to
gravity, salinity volumetric expansion coefficient and kinematic
viscosity respectively. The horizontal characteristic velocity scale can be
rewritten as,  
\begin{equation}
  \label{osca}
u_c\sim\sqrt{g\beta\Delta C\delta_c}\,,\;v_c\sim u_c\left(\delta_c/L\right),\; p_c\sim\rho u_c^2
\end{equation}
$u_c$ can hence be viewed as a free fall velocity over the
characteristic boundary layer thickness.  The vertical characteristic
velocity is $\delta_c/L$ times smaller than $u_c$ and the predominant
balance is between motion pressure and the horizontal velocities. The
motion pressure, $\tilde p\,=\,\tilde p_s-\tilde p_h$, where $\tilde
p_s$ is the local static pressure in the presence of density
variations, and $\tilde p_h$ is the local hydrostatic pressure when
the fluid through out is at the reference density $\tilde\rho_r$.

Normalising the governing equations of two dimensional continuity,
Navier-Stokes and species equation with these characteristic scales,
dropping terms of order $\leq 1/Gr_L^{2/5}$ when
$Gr_L\rightarrow\infty$, and assuming Schmidt number, $Sc\sim 1$ we
obtain the non-dimensional boundary layer equations as,
\begin{equation}\label{bcty}
\frac{\partial u }{\partial x}+\frac{\partial
v}{\partial y}=0 
\end{equation}
\begin{equation}\label{bxmom}
u\frac{\partial u}{\partial x}+v\frac{\partial u}{\partial
y}=-\frac{\partial p}{\partial x}+\frac{\partial ^2 u}{\partial y^2}
\end{equation}
\begin{equation}\label{bymom}
\frac{\partial p}{\partial y}= \,c
\end{equation}
\begin{equation}\label{bspceq}
u\frac{\partial c}{\partial x}+v\frac{\partial c}{\partial y}
=\frac{\partial ^2 c}{\partial y^2}
\end{equation}
with the boundary conditions,
\begin{equation}\label{bc1}
\textrm{at y =} \delta(x):\;\;\;u = 0,\quad\frac{\partial u}{\partial y}=0,\qquad
c=\,0,\qquad\frac{\partial c}{\partial y}=0
\end{equation}
\begin{equation}\label{bc2}
\textrm{at y = 0}:\quad u = 0,\quad c = 1\quad
v =\frac{\tilde v_w}{v_c}\,=\,\frac{Re_w}{Gr_L^{1/5}}\,=V_w,
\end{equation}
where $ x=\tilde x/L,\,y=\tilde y/\delta_c,\,u=\tilde u/u_c,\,v=\tilde
v/v_c,\,p=\tilde
p/p_c,\,c\,=\,\frac{\tilde{C_o}-\tilde{c}}{\Delta\tilde{ C}},\,
\delta(x)\,=\, \tilde\delta(\tilde
x)/\delta_c,\;Re_w=\tilde{v}_wL/\nu$.  Equation (\ref{bymom}) shows
that the vertical concentration distribution creates a vertical
pressure distribution, the horizontal gradient of which in
(\ref{bxmom}) drives the flow. The concentration distribution is, in
turn, a result of this motion as well as diffusion.

The {\bf\emph{Blowing parameter}} $V_w$ is the non-dimensional normal
velocity at the wall. The blowing parameter also shows the condition
for similarity of the boundary layers in the presence of a weak
blowing.  When the normal velocity at the wall has a dependence of
$\tilde v_w \,\sim\,\tilde x^{-2/5}$ the boundary condition become
independent of $\tilde x$ and a similarity solution can be obtained.
The normalised terms in the boundary layer equations are of order one.
Hence these approximate equations can be expected to hold till the
blowing parameter $\frac{Re_w}{Gr_L^{1/5}}\,\sim$~O(1). We restrict
our calculations up to $V_w\,=\,3$.
\section*{\underline{Integral Formulation}}
\label{sec:integral-formulation-1}
Consider a steady laminar indirect natural convection boundary layer
with a weak blowing that could be described by the boundary layer
equations~(\ref{bcty}) - (\ref{bc2}).
Integrating ~(\ref{bcty}) from the surface to $\delta(x)$ and applying
the boundary conditions, we get the non dimensional integral mass
conservation equation as
\begin{equation}\label{intcty}  
\frac{d}{d x} \int_0^{\delta(x)}
u\;dy+v_\delta-V_w=0 
\end{equation} 
where, $v_\delta =\tilde{v}_\delta/v_c $ is the non-dimensional
entrainment velocity at the edge of the boundary layer. The
non-dimensional integral x momentum equation is obtained by integrating
the equation obtained after multiplying ~(\ref{bcty}) by u and
subtracting ~(\ref{bxmom}).
\begin{equation}\label{intmom} 
\frac{d}{dx} \int_0^{\delta(x)} u^2\;
dy+\int_0^{\delta(x)}\frac{\partial p}{\partial x}\; dy
+\frac{\partial u}{\partial y}_{|_{y=0}}=0 
\end{equation}
This equation expresses a balance of horizontal momentum flux with
horizontal pressure gradient and wall shear stress. Integrating
~(\ref{bspceq}) from 0 to $\delta(x)$ and substituting for $v_\delta$
from ~(\ref{intcty}), the non-dimensional integral species equation is
obtained as
\begin{equation}\label{intspceq}
-\frac{d}{dx}\int_0^{\delta(x)} u c\; dy 
+\frac{\partial c}{\partial
y}_{|_{y=0}}\,+V_w\,=\,0
\end{equation} 
The terms in the above equation represent convection by the horizontal
velocity and entrainment from ambient, the diffusive flux at the wall
and the blowing flux at the wall respectively. The concentration
profile is approximated by the function,
\begin{equation}\label{cprof}
f(\eta,a)\,=\,c\,= {\left( 1 - \eta  \right) }^2\,
\left( 1 + a\,\eta  \right)
\end{equation}
where, $\eta\,=\,\tilde y/\tilde \delta(\tilde{x})$ and the
coefficient \emph{a} captures the non similar nature of the profile in
the presence of blowing.  The profile ~(\ref{cprof}) satisfies the
boundary conditions ~(\ref{bc1}) and ~(\ref{bc2}) ie.,
$f(0,a)=1,\:f(1,a)=0,\;\dot{f}(1,a)=0$, where,
superscript\textbf{\,\Huge $.$} denote differentiation with respect
to $\eta$.

The coefficient $a$ in~(\ref{cprof}) is found from the condition that
the profile satisfies the species equation ~(\ref{bspceq}) at the wall
ie,
\begin{equation}\label{acond}
\ddot{f}(0,a)=V_w\, \delta\dot{f}(0,a)
\end{equation}
Substituting the values of $\dot{f}(0,a)\,=\,a-2\,
\mathrm{and}\,\ddot{f}(0,a)\,=\,2-4a$ in ~(\ref{acond}) and solving
for $a$,we get
\begin{equation}\label{a} 
a(V_w,x)=2 - \frac{6}
{4 +V_w\, \delta} 
\end{equation} 
The velocity profile is  approximated by the polynomial
\begin{equation}\label{uprof}
u(\eta, {b})\,=\,c_1\,+{b}(x)\,\eta\,+\,c_2\,\eta^2\,
+\,c_3\,\eta^3\,+\,c_4\,\eta^5 
\end{equation} 
where the dimensionless coefficient \emph{b} captures the non similar
nature of the profile in the presence of blowing.  We need to satisfy
the boundary conditions ~(\ref{bc1}) and ~(\ref{bc2}), ie. $ u(0,
{b})\,=0,\:u(1, {b})\,=0,\:\textrm{and}\:\dot{u} (1, {b})\,=0,$ and
the condition obtained from the x-momentum equation at the wall, 
\begin{equation}
\frac{V_w}{
  \delta}\dot u(0)-\left( \delta\,
I_1\right)^\prime\,-\frac {\ddot u(0)}{ \delta^2}=0
\end{equation} 
where $^\prime$ denotes differentiation with respect to x.
In the above equation, the horizontal pressure gradient in
~(\ref{bxmom}) is expressed in terms of the concentration profile
using ~(\ref{bymom}) as
\begin{equation}
\left.\frac{dp}{dx}\right|_{\eta=0}\,=\,-(\delta I_1)^\prime\,\;\textrm{ where,}\,I_1\,=\,\int_0^1\,f(\eta)d\eta
\end{equation}

Using the above conditions, $c_1\,=\,0$, and $\,c_2,\,c_3,\,c_4\,$ can
be expressed as functions of $ b,\,V_w,\,\textrm{and}\, \delta$. The
final expressions of the coefficients are given in the Appendix.  The
velocity profile is obtained as \small
\begin{equation}
\raggedleft
u(\eta, b)\,=\,\eta\,(\eta-1)^2\left[\frac{ b
(x)}{2}\,\left(2+\eta(4+V_w\,
 \delta)\right)-\frac{\eta(2+V_w\,
 {\delta})(6+V_w\, \delta)}{4(4+V_w\, \delta)^2
}\, \delta^2 \delta^\prime\right] 
\end{equation} 
\normalsize 
Rewriting ~(\ref{intspceq}) and ~(\ref{intmom}) in terms
of the profiles ~(\ref{cprof}) and ~(\ref{uprof}), we get the integral
equations for species and momentum.
\begin{eqnarray}
\left[ \delta\, I_{uf}\right]^\prime-V_w+
\frac{\dot f(0,a)}{ \delta}=0\label{fispceq}\\
\left[\, \delta\, I_{u^2}\,- \delta^2\,
I_2\right]^\prime+\frac{\dot{u}(0, b)}{ \delta}
\,=\,0\label{fimom}
\end{eqnarray}
where,
\small
\begin{equation}
I_{uf}=\int_0^1u(\eta,x)\,f(\eta,a)\,d\eta,\; 
I_{u^2}= \int_0^1u(\eta,b)^2\,d\eta;\:I_2\,=\,\int_0^1\,\int_\eta^1f(\eta,a)\,d\eta\:d\eta
\end{equation} 
\normalsize
In ~(\ref{fimom}) above, the horizontal pressure gradient in
~(\ref{intmom}) has been replaced in terms of the concentration profile
by integrating the y momentum equation ~(\ref{bymom}) from
$1\,\textrm{to}\,\eta$ as below 
\begin{equation}
\int_0^{\delta(x)}\frac{\partial p}{\partial x}\;
  dy=\,-(\delta ^2\,I_2)^\prime \label{prg}
\end{equation}
Substituting the values
$ {I}_{uf},\, {I}_{u^2},\, {I}_2,\,\dot{u}(0),\,
\dot{ {f}}(0)$ and simplifying, we get two autonomous nonlinear
second order ordinary differential equations for the unknowns
$\delta(x)$ and $ b(x)$ as follows.
\begin{eqnarray}
 {\delta}^{\prime\prime}\,+\, {\delta}^\prime\left(A\, {\delta}^
\prime+\,B\right)+\,C\, {b}^\prime\,+\,D\,=\,0\label{fde1}\\
 {\delta}^{\prime\prime}\left(1+E {\delta}^\prime\right)\,+\, {\delta}^\prime\left(\,F\, {\delta}^{\prime^2}\,+\,G\, {\delta}^\prime\,+\,H\right)+\, {b}^\prime\left(I {\delta}^\prime\,+J\right)+\,K=0\label{fde2}
\end{eqnarray} 
where the coefficients, A to K are nonlinear functions of
$ {\delta}(x),\,V_w\,\textrm{and}\, {b}(x)$
The expressions for these coefficients are given in the Appendix.
 
Equations ~(\ref{fde1}) and ~(\ref{fde2}) needs three initial
conditions $ {\delta}(0),\, {\delta}^\prime(0),$ and $ {b}(0)$ for
solution as an initial value problem.  We assume that as $x\rightarrow
0$ the boundary layer behaves similar to the case with no blowing.
This is equivalent to assuming an infinitesimal distance where no
blowing effects are present. Hence, we use the values of $
{\delta}(\epsilon)\,=\,0.0179,\, {\delta}^\prime(\epsilon)\,
=\,6369.7$ and $ {b}(\epsilon)\,=\,0.2358$, where $\epsilon =10^{-6}$,
as calculated from the similarity solutions of Rotem and
Classen~\citep{rc}as the initial conditions.  It was found that the
solution was insensitive to changes in initial conditions up to an
order magnitude from the above conditions. The NDSolve routine of
MATHEMATICA was used to numerically solve ~(\ref{fde1}) and
~(\ref{fde2}) simultaneously.
\section*{\underline{Results and Discussion}}
\label{sec:results-discussion}
\begin{figure}[!tbp]
\centering
  \subfigure[]{\label{rcd}
    \includegraphics[width=0.32\textwidth]{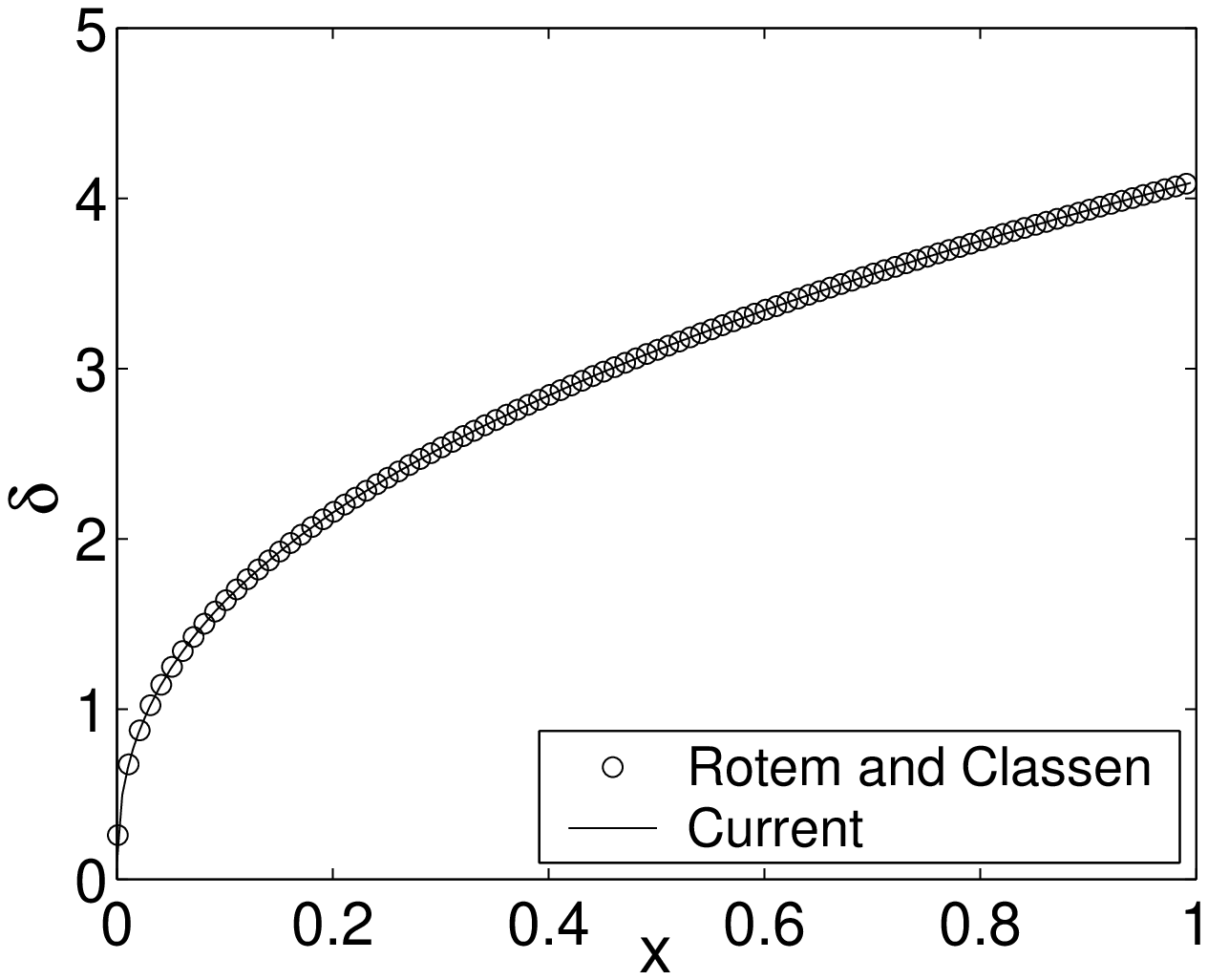}}
  \subfigure[]{\label{rcf}
    \includegraphics[width=0.32\textwidth]{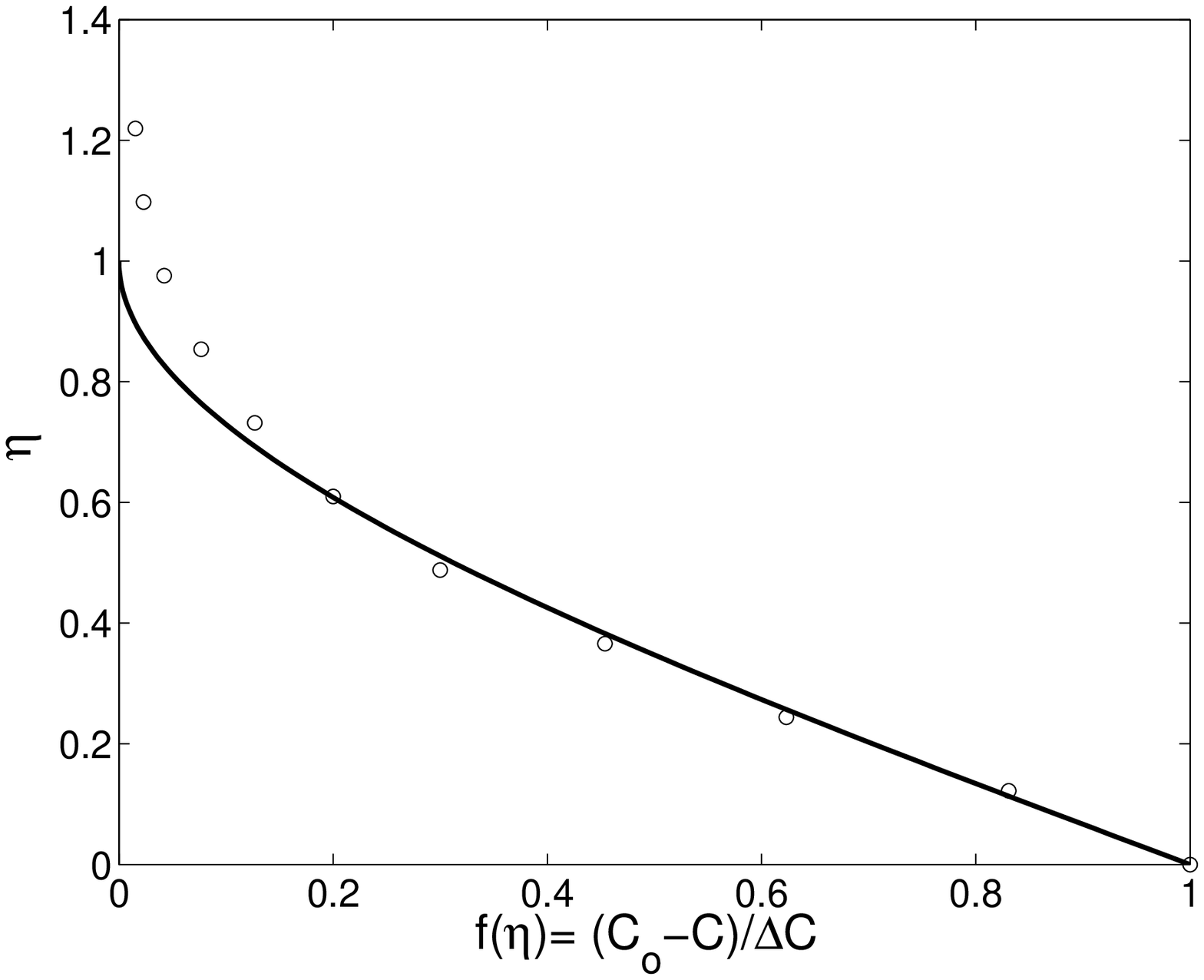}}
  \subfigure[]{\label{rcu}
    \includegraphics[width=0.32\textwidth]{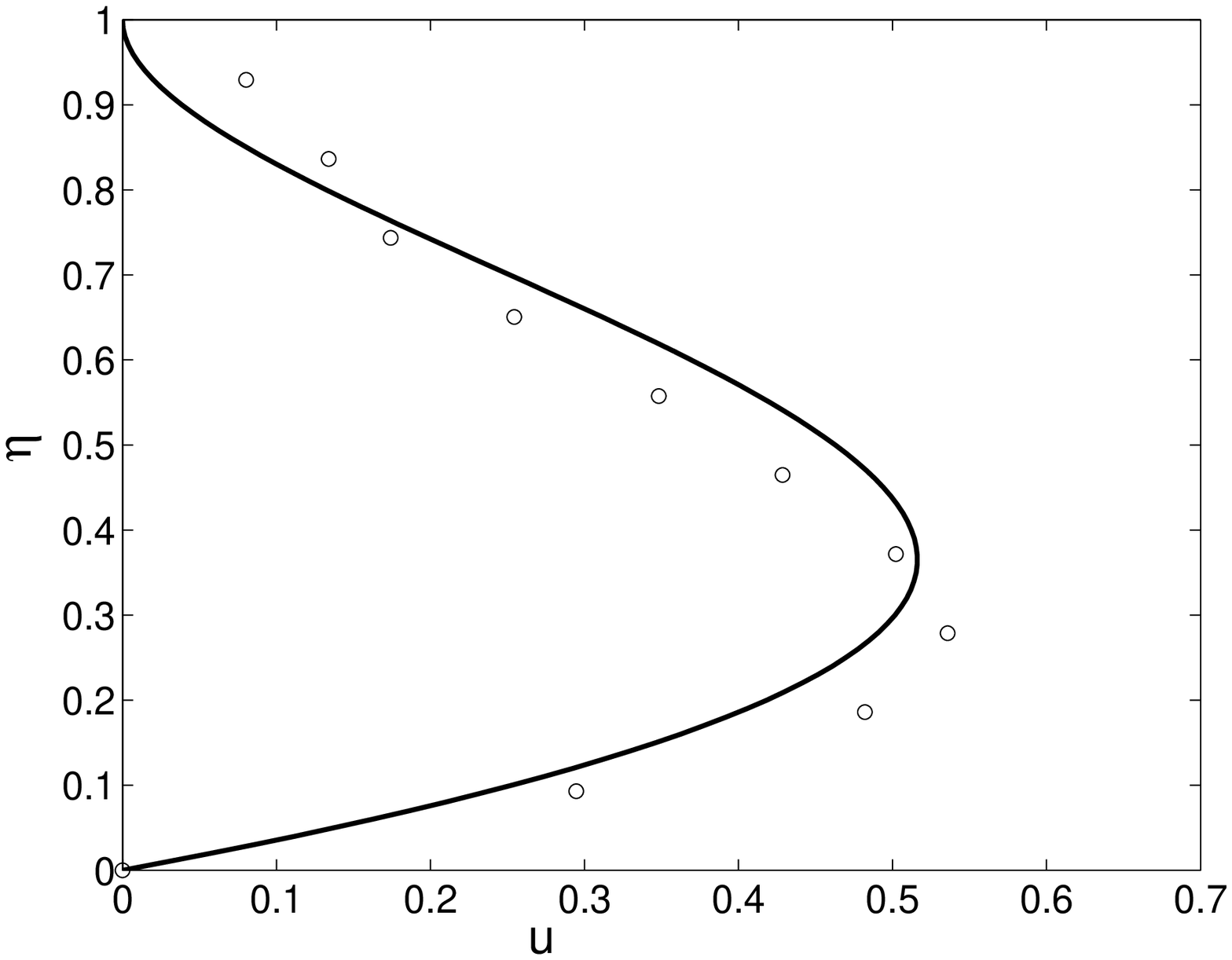}}
\caption{Comparison with Rotem and Classen~\citep{rc} for no blowing case. (a) Non dimensional boundary layer thickness  (b) Non-dimensional concentration profile  (c) Non-dimensional Velocity profile }
\end{figure}

The results are initially compared with similarity solutions for the
no blowing case by Rotem and Classen~\citep{rc}. The integral analysis
gives ${\delta}(x)\,\sim \,x^{2/5}$, same as that obtained in the
similarity solution. The boundary layer thickness $\delta$ in the
integral analysis is equal to the thickness at which c = 0.05 in the
solution of Rotem and Classen ~\citep{rc}.  Figure~(\ref{rcd})
compares the $x$ dependence of the non dimensional boundary layer
thickness $\delta$ with that of Rotem and Classen~\citep{rc}.Figures
(\ref{rcf}) and ~(\ref{rcu}) compare the concentration and velocity
profiles obtained from the integral analysis with those from the
similarity solution. Considering the approximate nature of integral
methods, the comparison is satisfactory.

\begin{figure}[!tbp]
  \subfigure[]{\label{dtf} \includegraphics[width=0.45\textwidth]{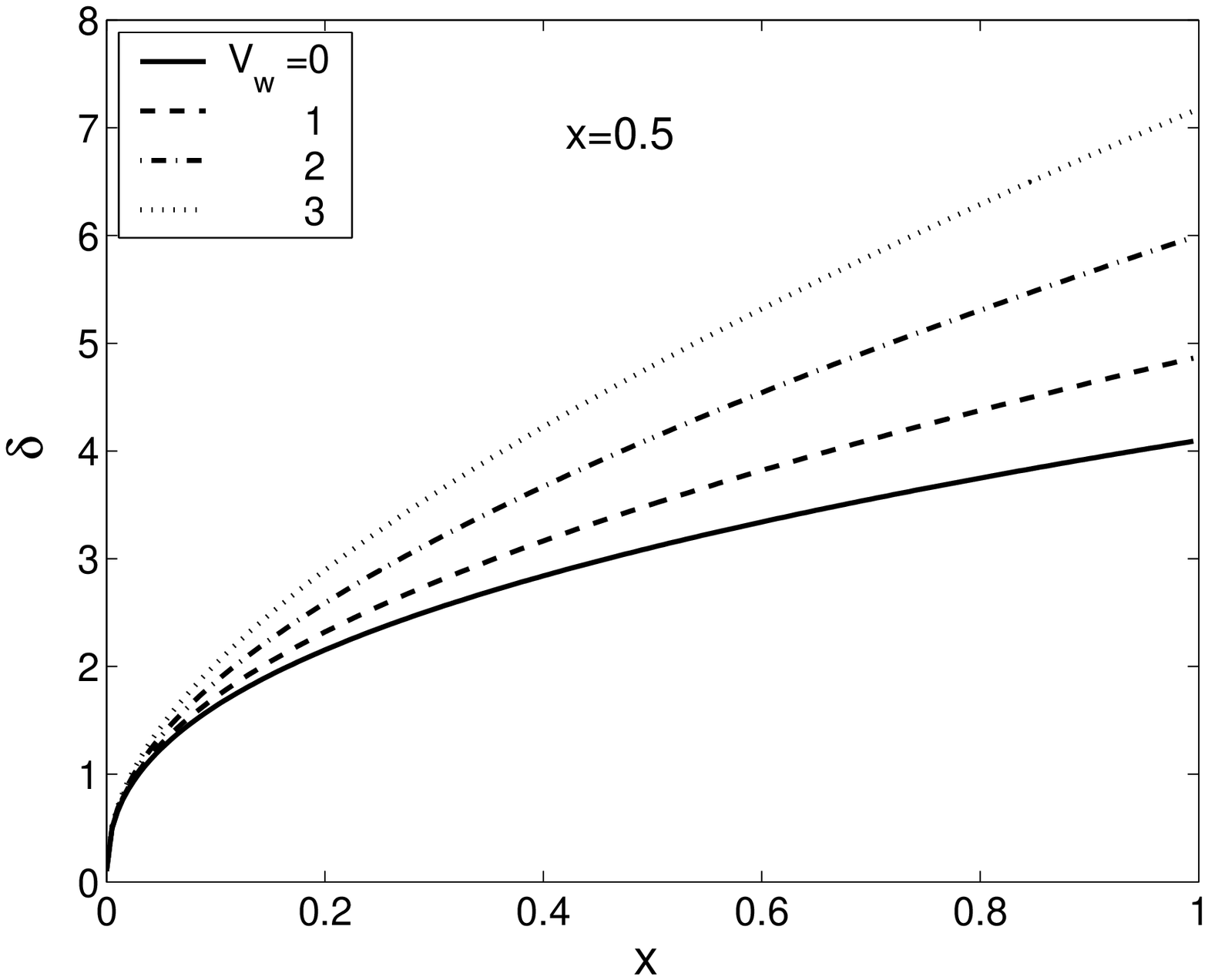}}\hfill
  \subfigure[]{\label{ftf}\includegraphics[width=0.45\textwidth]{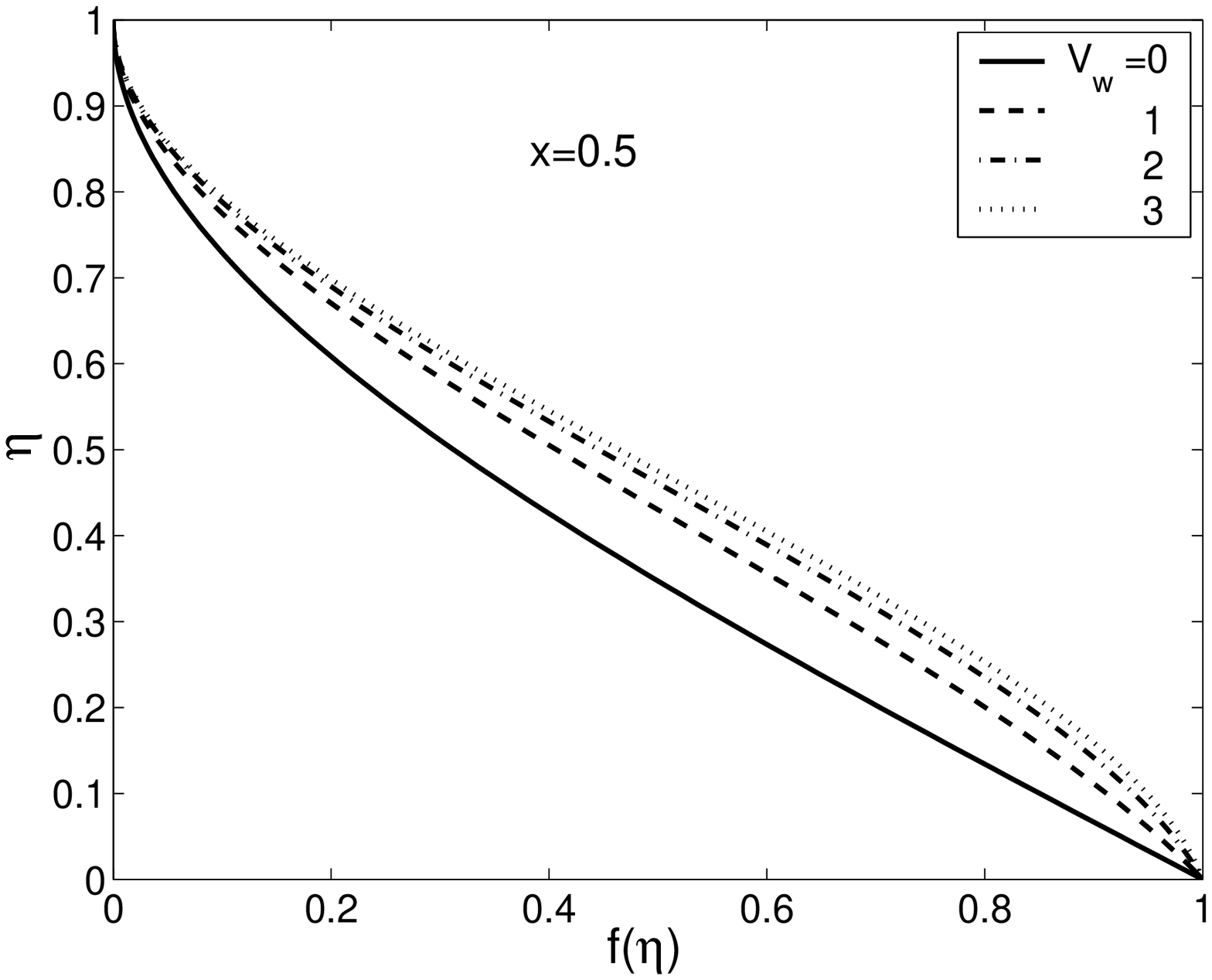}}\hfill
\caption{Effect of blowing on (a) boundary layer thickness and (b) concentration profile  }  
\end{figure}

Figure~(\ref{dtf}) shows the effect of blowing on the boundary layer
thickness. As expected, higher blowing results in larger boundary
layer thickness; the increase in total added fluid increases the
boundary layer thickness.
The concentration profiles at $x= 0.5$ for different blowing
parameters are shown in Figure~(\ref{ftf}).  The reduction in the
gradient of concentration at the wall shows that the diffusive flux
decreases with increase in the blowing parameter. 

\begin{figure}[tbp]
  \subfigure[]{\label{utf}\includegraphics[width=0.45\textwidth]{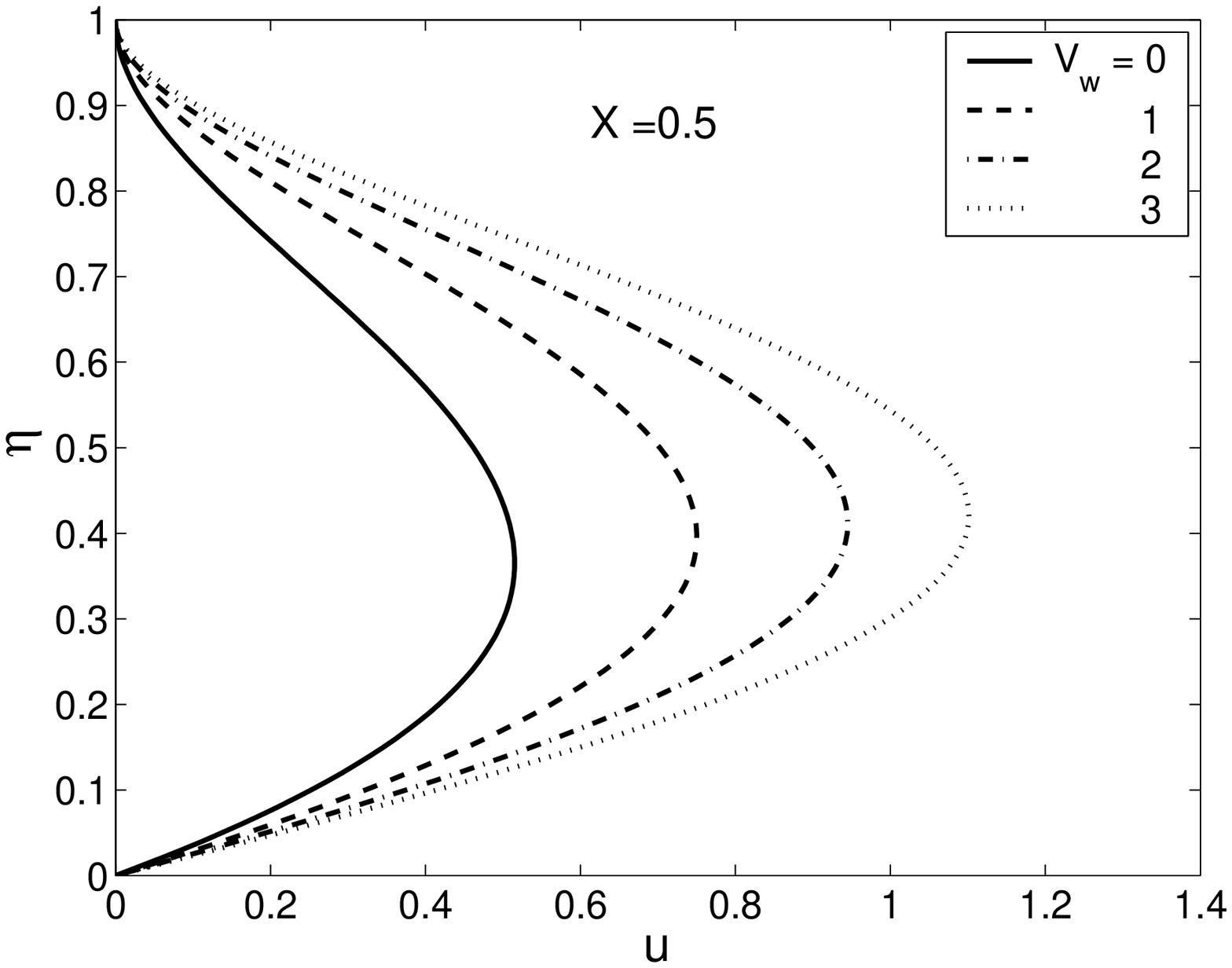}} \hfill
  \subfigure[]{\label{fig:pg}\includegraphics[width=0.45\textwidth]{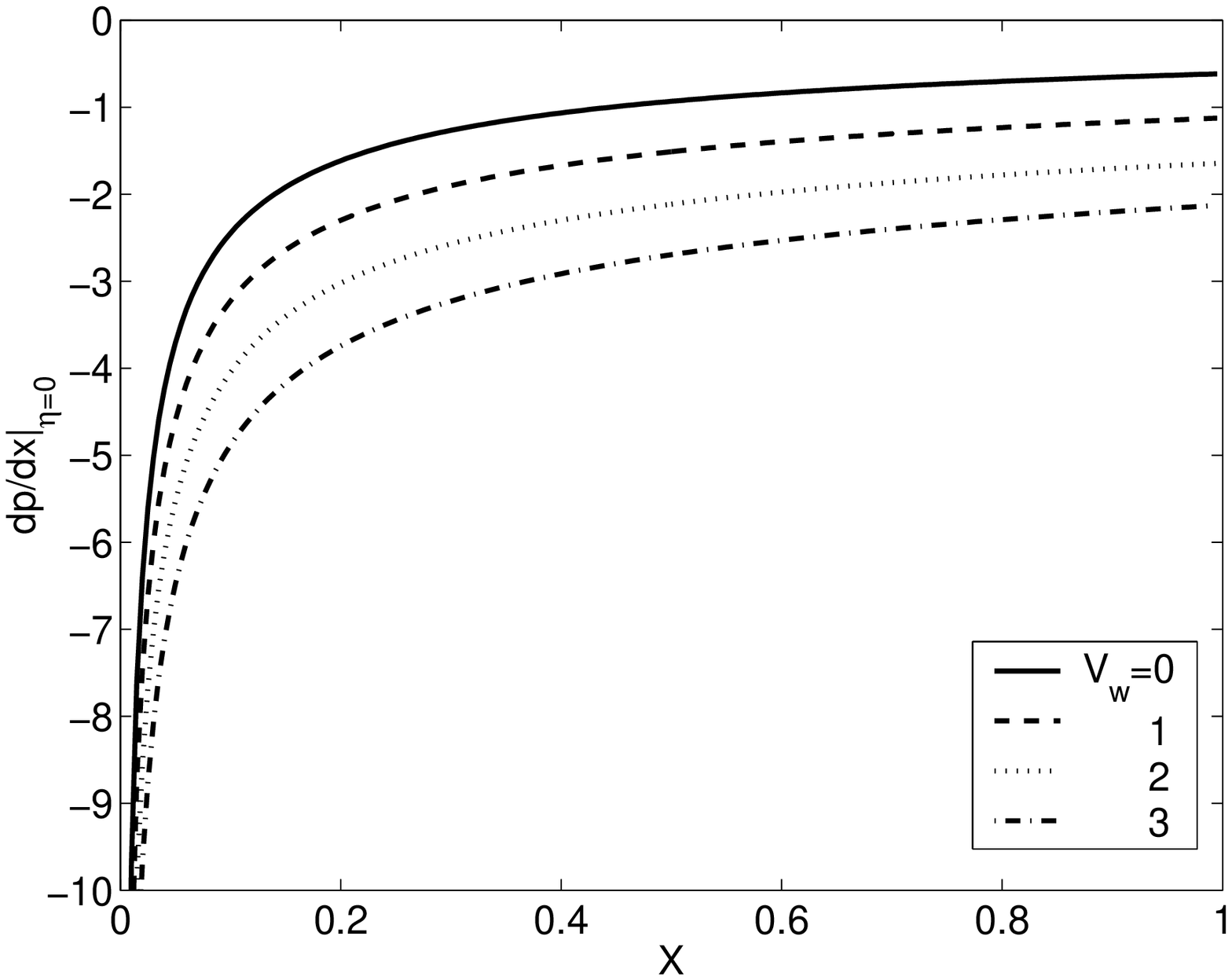}}
  \caption{Effect of blowing on (a)horizontal velocity profile and (b)horizontal pressure gradient at the wall}
\end{figure}

The major effect of the change in concentration profile shape is on
the pressure distribution. The motion pressure outside the boundary
layer is zero. The motion pressure within the boundary layer is given
by $\tilde p(\tilde y,\tilde x)=-\rho g\beta\int_{\tilde y}^{\tilde
  y=\tilde\delta}(\tilde C_o-\tilde c)\, d\tilde y$. Presence of
increased amounts lighter fluid in the boundary layer due to blowing
results in larger value of $\tilde C_o-\tilde c$. This results in a
larger favourable horizontal pressure difference with the outside
motion pressure than in the case of without blowing, resulting in
larger velocities. This could be observed in Figure~(\ref{utf}) shows
the non dimensional horizontal velocity distribution for different
blowing parameters at x=0.5. Figure~\ref{fig:pg} shows the variation
of horizontal pressure gradient at the wall as a function of
horizontal distance for increasing value of the blowing parameter.
The pressure gradient was calculated using the relation $\frac{d\tilde
  p}{d\tilde x}= -\rho g\beta\frac{d}{d\tilde x}\int_{\tilde
  y=0}^{\tilde y~=~\tilde \delta}(\tilde C_o-\tilde c)d\tilde y$. The
increased amounts of lighter fluid in the boundary layer has the
effect of making the horizontal pressure gradient at the wall more
favourable. This increases $\frac{d\tilde u }{d\tilde y}$ at the wall
and results in increased wall shear stresses with blowing.  This
behaviour is in contrast to what is observed in a shear boundary layer
with blowing.  For example, in the case of the Blasius boundary layer,
blowing reduces the wall shear stress.
The vertical characteristic velocity in the boundary layer is
$\frac{\delta_c}{L}$ times the horizontal characteristic velocity, as
can be seen from Equation~(\ref{osca}). Therefore, the vertical
momentum from blowing velocities of the order of
$v_c\:ie,\:V_w\,\sim\,O(1)$, doesn't seem to be sufficient to cause a
change in the near wall linearity of the horizontal velocity profile.
For the current range of blowing parameters investigated, the major
effect seems to be due to increase in motion pressure, which increases
the horizontal velocities, and the wall shear stress.

The total mass flux of the species into the boundary layer is
\begin{equation}\label{dq} 
\tilde q\,=\,\tilde v_w\,\Delta \tilde{C}\, +\,D\frac{\partial \tilde c}
{\partial \tilde y}|_{\tilde y=0} 
\end{equation} 
The first term on the right hand side of ~(\ref{dq}) is due to
blowing(advection) and the second term represents diffusion. The
non-dimensional flux is,
\begin{equation}\label{ndq}  
 q\,= \,\frac{\tilde q}{D \frac{\Delta C
}{\delta_c}}\,=\,V_w\,-\,\frac{f^\prime(0)}{ {\delta}}
\end{equation} 

\begin{figure}[!tbp]
\parbox[l]{0.5\textwidth}{
\centering
\includegraphics[width=0.45\textwidth]{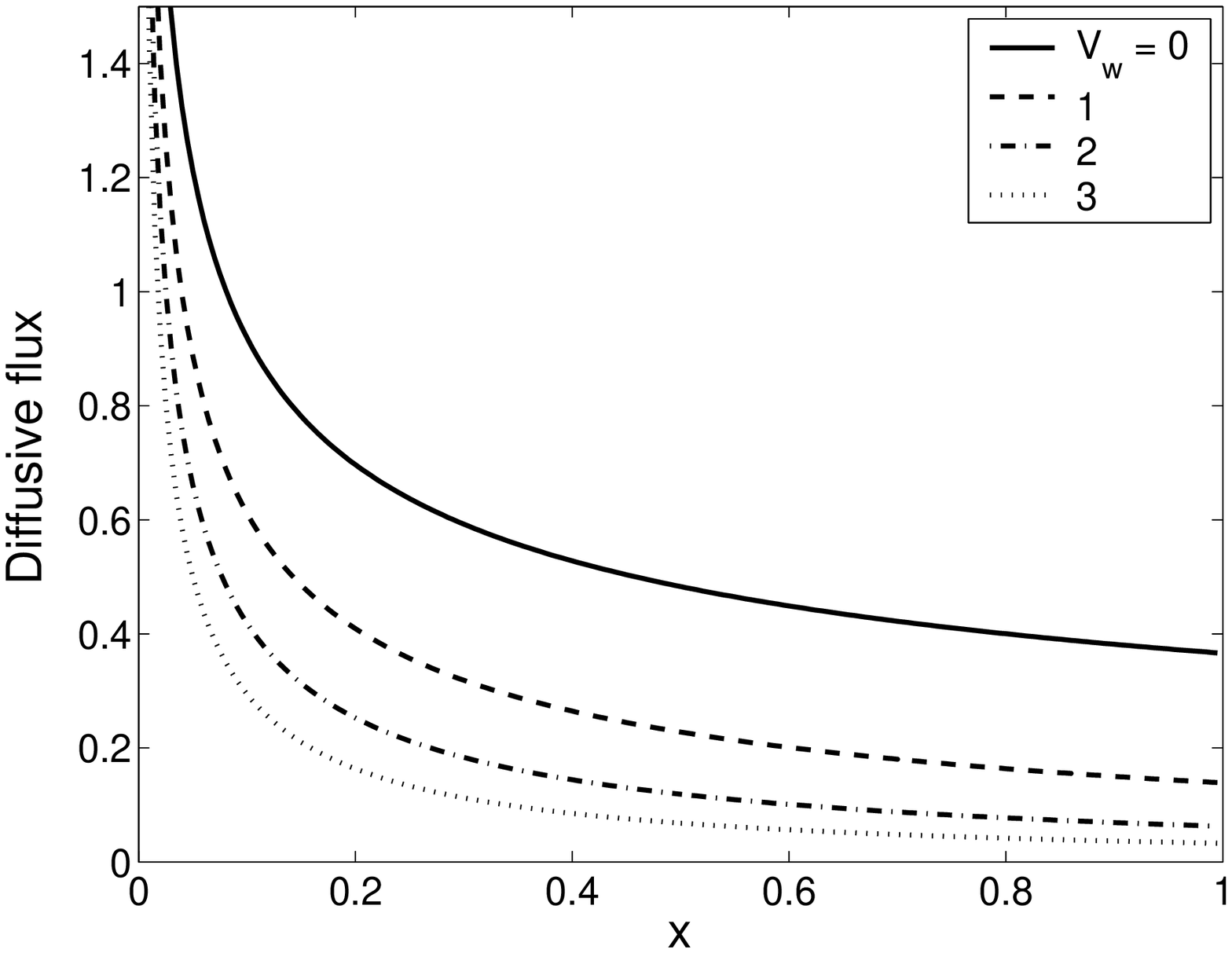}
\label{fig:qdx}
\caption{Variation of the longitudinal distribution of diffusive flux with blowing parameter}}
\parbox[r]{0.5\textwidth}{
\centering
  \includegraphics[width=0.45\textwidth]{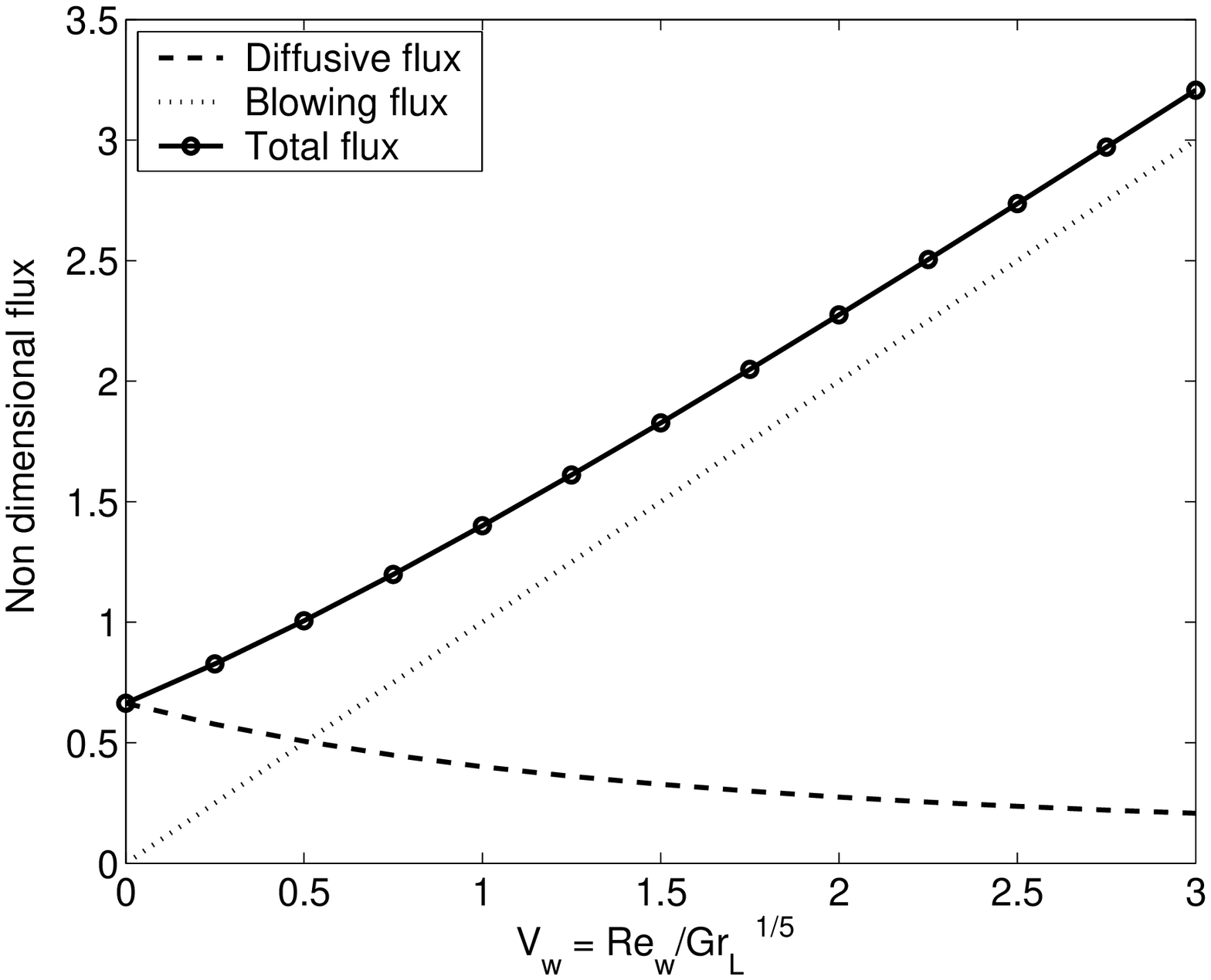}
\caption{Variation of non-dimensional fluxes with blowing parameter}\label{qvw}}
\end{figure}

The variation of non-dimensional diffusive flux over the length of the
boundary layer is shown in Figure(5)~\ref{fig:qdx}. Blowing reduces the
diffusive flux.  Note that when $V_w\,>\,1$, except very near the
leading edge, the diffusive flux is an order lower than the blowing
flux, which is equal to $V_w$. 

Figure~(\ref{qvw}) shows the longitudinally averaged blowing and
diffusive fluxes as a function of $V_w$. The total flux is also shown
in the figure. As pointed above, the mean diffusive flux drops with
$V_w$. At $V_w\,=\,3,$ the diffusive flux is about $\frac{1}{3}$rd the
diffusive flux at no blowing.  The blowing flux increases linearly
with the blowing parameter, and at $V_w\,=\,3,$ the blowing flux is
about fifteen times the diffusive flux, and about 5 times the no
blowing flux.  The variation of the ratio of the total flux with
blowing to the flux without blowing
($\frac{q}{\frac{f^\prime(0)}{\delta}}$) could also be noticed from
Figure~(\ref{qvw}).  For example, the flux with $V_w$=1 is double the
flux with no blowing.  To take a specific example, this means that at
a commonly encountered $Gr_L=10^6$ in water, blowing velocities of the
order of 1 mm/s doubles the flux from the no blowing case. Hence, weak
blowing increases the flux appreciably over the corresponding no
blowing flux.
\section*{\underline{Conclusions}}
\label{sec:conclusions}
The  integral analysis developed in  this paper is   used to study the
effects of   weak   blowing on   indirect laminar natural   convection
boundary  layers.   The blowing velocity   and  wall concentration are
constant along  the wall.  The analysis is  subject to the assumptions
of small $\delta/L$(large Grashoff numbers), unity Schmidt(or Prandtl)
numbers  and  blowing  velocities of the   same order  as the vertical
characteristic velocities   due  to free convection   in  the boundary
layer.  Approximate solutions were obtained by numerically solving the
integrated  boundary   layer  equations.    Following are    the major
conclusions from  the  study: (a)   As expected,   the boundary  layer
thickness increases with blowing  (Figure~\ref{dtf}). (b) Weak blowing
increases the horizontal velocities  in the boundary  layer, resulting
in      the  wall  shear       stress   also increasing   with blowing
(Figure~\ref{utf}). (c) The  concentration  gradient at the   wall and
thus the diffusive flux reduces with increase in the blowing parameter
(Figure~\ref{ftf}).   (d) For  $V_w >  1$,  the flux  due to advection
caused by  blowing becomes the  predominant contribution  to the total
flux and hence the total flux shows a  linear dependence on $V_w$. (e)
Comparison of total flux with that in the case of no blowing show that
a   weak    blowing   results  in  considerable  increase     in  flux
(Figure~\ref{qvw}).
\section*{\underline{Appendix }}
\label{sec:appendix}
The coefficients in the velocity profile ~(\ref{uprof}) are,
\begin{eqnarray}
c2=  5\, {\delta} \, \,  {b}  -\,Q\,  {\delta}^2
\, {\delta}^\prime,\quad  
c3=-  {b} \,\left( 3 + 10\, {\delta}  \right)    +2\,Q\,
 {\delta} ^2\, {\delta}^\prime, \quad
c4=\frac{  {b} \, \left( 4 + 10\, {\delta}  \right) }{2} -\,Q\,
 {\delta} ^2\, {\delta}^\prime\\ 
\textrm{where},\:Q\,=\,\frac{\left( 2 + 10\, {\delta}
\right) \, \left( 6 + 10\, {\delta}  \right) \, }{4\, {\left( 4 +
10\, {\delta}  \right) }^2}
\end{eqnarray}
The coefficients in the differential equations ~(\ref{fde1}) and
~(\ref{fde2}) are obtained as
\begin{eqnarray}
A=\frac{3}{\delta } + V_w\,\left( \frac{1} {2 +
V_w\, {\delta} } - \frac{3}{4 + V_w\, {\delta} } +
\frac{1}{6 + V_w\, {\delta} } + \frac{7}{19 +
7\,V_w\, {\delta} } \right)\\ 
B=\frac{-2\, {b}\, \left( 4 + V_w\, {\delta}  \right) \,
\left( 816 + 7\,V_w\, {\delta} \, \left( 104 +
V_w\, {\delta} \, \left( 25 + 2\,V_w\, {\delta}
\right) \right)  \right) }{{ {\delta} }^3\, \left( 2 +
V_w\, {\delta}  \right) \, \left( 6 + V_w\, {\delta}
\right) \, \left( 19 + 7\,V_w\, {\delta}  \right) }\\
C=\frac{-2\,{\left( 4 + V_w\, {\delta}  \right) }^2\, \left( 204
+ 7\,V_w\, {\delta} \, \left( 13 + V_w\, {\delta}
\right)  \right) }{{ {\delta} }^2\, \left( 2 +
V_w\, {\delta}  \right) \, \left( 6 + V_w\, {\delta}
\right) \, \left( 19 + 7\,V_w\, {\delta}  \right) },\quad
D=\frac{1680\,{\left( 4 + V_w\, {\delta}  \right) }^ 2\,\left( 6
+ V_w\, {\delta} \, \left( 4 + V_w\, {\delta}  \right)
\right) } {{ {\delta} }^4\, \left( 2 + V_w\, {\delta}
\right) \, \left( 6 + V_w\, {\delta}  \right) \, \left( 19 +
7\,V_w\, {\delta}  \right) }\\
E=-\frac{{ {\delta} }^2\, \left( 2 + V_w\, {\delta}
\right) \, \left( 6 + V_w\, {\delta}  \right) }{  {b}\,
{\left( 4 + V_w\, {\delta}  \right) }^2\, \left( 17 +
2\,V_w\, {\delta}  \right) },\quad 
F=-\frac{-\delta \,
      \left( 240 + V_w\, {\delta}\,
         \left( 236 + 5\,V_w\, {\delta}\,
            \left( 12 + V_w\, {\delta} \right) 
           \right)  \right)  }{2\, {b}\,
    {\left( 4 + V_w\, {\delta} \right) }^3\,
    \left( 17 + 2\,V_w\, {\delta}\right) }\\
G= \frac{3}{ {\delta}} + 
  V_w\,\left( \frac{1}
      {2 + V_w\, {\delta}} - 
     \frac{2}{4 + V_w\, {\delta}} + 
     \frac{1}{6 + V_w\, {\delta}} + 
     \frac{2}{17 + 2\,V_w\, {\delta}}
\right)
\end{eqnarray}
\begin{eqnarray}
H= \frac{504\, {\delta}\,
     \left( 2 + V_w\, {\delta} \right) \,
     \left( 16 + 3\,V_w\, {\delta} \right)}{{ b\,
     {\delta}^3\,
    \left( 2 + V_w\, {\delta} \right) \,
    \left( 6 + V_w\, {\delta} \right) \,
    \left( 17 + 2\,V_w\, {\delta} \right) }} \nonumber\\ 
-\frac{ 2\, {b}^2\,\left( 4 + V_w\, {\delta} \right)^2\,
     \left( 76 + V_w\, {\delta}\,
        \left( 34 + 3\,V_w\, {\delta} \right) 
       \right) }{{ b\,
     {\delta}^3\,
    \left( 2 + V_w\, {\delta} \right) \,
    \left( 6 + V_w\, {\delta} \right) \,
    \left( 17 + 2\,V_w\, {\delta} \right) }}
\end{eqnarray}
\begin{eqnarray}
I= \frac{1}{ {b}},\quad 
J=\frac{-4\,{\left( 4 + V_w\, {\delta} \right) }^2\, \left( 76
+ V_w\, {\delta} \, \left( 17 + V_w\, {\delta}
\right)  \right) }{{ {\delta} }^2\, \left( 2 +
V_w\, {\delta}  \right) \, \left( 6 + V_w\, {\delta}
\right) \, \left( 17 + 2\,V_w\, {\delta}  \right) }\\
K=\frac{-5040\,{\left( 4 + V_w\, {\delta}  \right) }^
2}{{ {\delta} }^4\, \left( 2 + V_w\, {\delta}  \right) \,
\left( 6 + V_w\, {\delta}  \right) \, \left( 17 +
2\,V_w\, {\delta}  \right) }
\end{eqnarray}
\bibliography{babu}

\begin{thebibliography}{1}

\bibitem{geb}
{B.Gebhart}{et.al}.
\newblock \textit{Buoyancy induced flows and transport}.
\newblock Hemisphere Publishing (1988).

\bibitem{blt}
H.~Schlichting and K.~Gersten.
\newblock \textit{Boundary layer theory}.
\newblock Springer-Verlag (2000).

\bibitem{gzc}
D.~Gill, W.N.~Zeh and E.~Casal.
\newblock Free convection on a horizontal plate.
\newblock \textit{Z. Angew. Math. Phys.} Vol 16(1965), pp 539.

\bibitem{rc}
Z.~Rotem and L.~Claassen.
\newblock Natural convection above unconfined horizontal surfaces.
\newblock \textit{Jl. Fluid. Mech.} 39, part1(1969), pp 173.

\bibitem{cr}
J.~Clarke and N.~Riley.
\newblock Natural convection induced in a gas by the presence of a hot porous
  horizontal surface.
\newblock \textit{Ql. Jl. Mech. appl. Math.} XXVIII, Pt.4(1975), pp 375.

\end{thebibliography}
\end{document}